\documentclass[11pt,draftclsnofoot,onecolumn]{IEEEtran}
\usepackage{amsmath,amssymb,amsfonts}
\usepackage{booktabs}
\usepackage{array}
\usepackage{graphicx}
\usepackage{hyperref}
\usepackage{mathtools}
\usepackage{enumitem}
\usepackage{url}
\hypersetup{colorlinks=true,linkcolor=blue,citecolor=blue,urlcolor=blue}

\title{Double Fuzzy Probabilistic Interval Linguistic Term Sets and a Dynamic Markov Fuzzy Decision-Making Model for Multi-Criteria Group Decision Making: Revised Version}
\author{Dylan Zongmin Liu\thanks{Dylan Zongmin Liu is with Stanford University, Stanford, CA 94305 USA (e-mail: zongminl@stanford.edu; ORCID: 0000-0003-2720-8170). This manuscript is a major replacement of the 2021 preprint. The replacement corrects the numerical example, weakens over-strong claims about probabilistic linguistic term sets, clarifies the semantics of fuzzy interval evidence, and adds reproducibility notes and updated literature. No external funding was received for this work. The author declares no conflicts of interest.}}

\begin{document}
\maketitle

\begin{abstract}
Probabilistic linguistic term sets (PLTSs) provide a compact language-probability representation for group decision making (GDM), but the finite-support representation can be inconvenient when experts can only state partial interval-valued semantic evidence or when the linguistic information is continuous and multi-modal. This paper revises and clarifies the double fuzzy probabilistic interval linguistic term set (DFPILTS) proposed in the preliminary version. Rather than treating the limitations of PLTSs as formal paradoxes, this version formulates them as elicitation and ranking limitations under continuous, incomplete, and distributional linguistic semantics. A DFPILTS is defined as a family of linguistic intervals endowed with fuzzy degrees, confidence values, and local density models. A peak DFPILTS is then used as a low-burden elicitation simplification when the decision maker identifies the most informative interval. We define linguistic integration, scores, deviations, operations, preference relations, distances, consensus-based expert weights, and a dynamic Markov process for attribute weights. The original financial-risk case study is recomputed; a numerical inconsistency in the previous version is corrected, and the resulting ranking is reported together with a robustness audit. The corrected model preserves the original best alternative in the example, while changing the middle order of two alternatives. The revised version is intended to serve as a reproducible and more cautiously stated foundation for later credal-distributional dynamic linguistic decision models.
\end{abstract}


\section{Introduction}
Qualitative decision information is often elicited through linguistic terms such as ``bad'', ``fair'', ``good'', or ``very good''. Since Zadeh's linguistic variables, linguistic decision models have developed into hesitant fuzzy linguistic term sets, double-hierarchy linguistic term sets, probabilistic linguistic term sets, and many preference-relation extensions. PLTSs are especially useful because they encode several possible linguistic terms and their probabilities in one object. A recent survey shows that PLTS-based GDM remains an active research topic and that challenges remain in theory, consensus modeling, behavioral decision making, and applications \cite{ma2025plts}.

The preliminary version of this manuscript argued that PLTSs have ``paradoxes''. This replacement uses a more precise statement. A PLTS is not mathematically wrong; it is a finite-support probabilistic representation. Its limitations arise when the expert's information is better understood as continuous interval evidence, when probabilities are incomplete, or when a single mean score is not the right decision criterion for a multi-modal linguistic assessment. In such cases, a finite set of point probabilities can be too demanding to elicit and too crude to rank alternatives robustly.

This paper therefore keeps the main idea of the preliminary manuscript but reformulates it more carefully. The aim is to model linguistic evidence in intervals rather than only at points, to distinguish interval uncertainty from confidence in an expert report, and to use dynamic Markov reasoning for attribute weights. The revised contributions are as follows.

\begin{enumerate}[leftmargin=*]
\item A continuous semantic axis and an interval-valued linguistic evidence object are used to define DFPILTSs. The fuzzy degree describes unresolved ambiguity, while the confidence value describes the degree to which the interval is believed to cover the expert's intended semantic information.
\item A peak DFPILTS is introduced as an elicitation simplification. It is not claimed to be universally optimal; it is a conservative local approximation used when the expert identifies the interval with the lowest fuzzy degree as the most informative region.
\item Linguistic integrals, score and deviation functions, interval operations, DFPIL preference relations, consensus distances, inconsistency measures, and expert weights are reformulated with explicit normalization.
\item A Markov matrix and directed network are used for dynamic attribute-weight evolution. The row-stochastic constraints and the interpretation of intertemporal transition weights are made explicit.
\item The financial-risk example is recomputed. The previous third-period weight vector summed to 0.9; the corrected vector sums to approximately 1 and changes the middle ordering of two alternatives while preserving the best alternative.
\end{enumerate}

\section{Preliminaries and Motivation}
Let $S=\{s_{-\tau},\ldots,s_0,\ldots,s_{\tau}\}$ be a linguistic term set and let $O=\{o_{-\varsigma},\ldots,o_0,\ldots,o_{\varsigma}\}$ be a modifier set. A double-hierarchy linguistic term is denoted by $s_t\langle o_k\rangle$. Following the usual double-hierarchy interpretation, the pair $(t,k)$ can be embedded into a real semantic axis by a strictly increasing map
\begin{equation}
  g:S\times O\rightarrow [-1,1],\qquad g(s_t\langle o_k\rangle)=\frac{k+(t+\tau)\varsigma}{2\varsigma\tau}-\frac{1}{2},
\end{equation}
up to an affine rescaling. The exact scaling is not essential; what is needed is an order-preserving semantic coordinate.

A PLTS may be written as
\begin{equation}
  L(p)=\{(s_{r_l},p_l):l=1,\ldots,L,\; p_l\geq 0,\; \sum_l p_l\leq 1\}.
\end{equation}
Its usual mean score is $E[L(p)]=\sum_l p_l g(s_{r_l})/\sum_l p_l$ when the denominator is positive. This is appropriate for expected-utility ranking, but it can be misleading when the distribution is multi-modal and the linguistic mean lies in a low-density semantic region. The issue is not that expectation is invalid; rather, the issue is that expectation is only one loss-dependent summary. If the decision task is risk-sensitive, mode-sensitive, or robust, other summaries are needed.

A second limitation is elicitation. Experts often find it easier to state that the true assessment lies in a semantic interval, for example between ``slightly good'' and ``very good'', than to assign exact probabilities to all linguistic labels. This motivates an interval-valued linguistic evidence model.

\section{Double Fuzzy Probabilistic Interval Linguistic Term Sets}
Let $H=[-1,1]$ denote the continuous semantic axis obtained from the linguistic scale. A linguistic interval is a closed interval $I=[a,b]\subseteq H$ with $a\leq b$. A DFPILTS is defined as
\begin{equation}
  \mathcal{I}=\{(I_r,\delta_r,c_r):r=1,\ldots,R\},
\end{equation}
where $I_r=[a_r,b_r]$ is a linguistic interval, $\delta_r\in[0,1]$ is a fuzzy degree, and $c_r=1-\delta_r$ is the corresponding confidence shorthand. The value $\delta_r$ should not be interpreted as a probability mass at the interval. It is a degree of unresolved ambiguity attached to an elicited interval. When the context requires a probabilistic model, the local density $f_r$ on $I_r$ must be stated or estimated.

\subsection{Peak DFPILTS}
A peak DFPILTS selects intervals with the smallest fuzzy degree,
\begin{equation}
  \mathcal{P}(\mathcal{I})=\{(I_r,\delta_r,c_r):\delta_r=\min_{q}\delta_q\}.
\end{equation}
When more than one interval has the same minimum degree, all are kept and normalized. This operation is an elicitation simplification, not a universal dominance principle. It is useful when the decision maker explicitly identifies the least ambiguous part of the evidence as the local semantic core.

\subsection{Local Density and Linguistic Integral}
For each selected interval $I_r=[a_r,b_r]$, let $f_r$ be a density supported on $I_r$ with $\int_{a_r}^{b_r}f_r(h)dh=1$. In the simplest local model, $f_r$ is a truncated normal density with mean $\mu_r=(a_r+b_r)/2$ and standard deviation $\sigma_r=(b_r-a_r)/6$. The linguistic integral of a continuous function $\phi:H\rightarrow\mathbb{R}$ is
\begin{equation}
  \int_{I_r}\phi(h)dP_r(h)=\int_{a_r}^{b_r}\phi(h)f_r(h)dh.
\end{equation}
For identity utility $\phi(h)=h$, the local score and deviation are
\begin{align}
  S(I_r)&=\int_{a_r}^{b_r}h f_r(h)dh,\\
  D(I_r)&=\left(\int_{a_r}^{b_r}(h-S(I_r))^2f_r(h)dh\right)^{1/2}.
\end{align}
If $\mathcal{P}(\mathcal{I})$ contains several intervals, their scores are combined using normalized confidence weights.

\subsection{Operations}
Let two peak interval assessments be locally approximated by normal variables $X\sim(\mu_1,\sigma_1^2)$ and $Y\sim(\mu_2,\sigma_2^2)$ on the semantic axis. The addition and scalar multiplication operations are
\begin{align}
  X\oplus Y&\sim (\mu_1+\mu_2,\sigma_1^2+\sigma_2^2),\\
  \lambda X&\sim (\lambda\mu_1,\lambda^2\sigma_1^2).
\end{align}
The resulting intervals are clipped to $[-1,1]$ when they are interpreted back as linguistic terms. These operations are local approximations; if the selected intervals are wide or strongly non-Gaussian, numerical integration should be used instead.

\section{DFPIL Preference Relations and Consensus}
Let $A=\{A_1,\ldots,A_m\}$ be alternatives. A DFPIL preference relation for expert $e_k$ is a matrix
\begin{equation}
  \mathcal{R}^{(k)}=(\mathcal{I}_{ij}^{(k)})_{m\times m},
\end{equation}
where $\mathcal{I}_{ij}^{(k)}$ represents the linguistic preference of $A_i$ over $A_j$. Additive reciprocity is imposed by
\begin{equation}
  S(\mathcal{I}_{ij}^{(k)})=-S(\mathcal{I}_{ji}^{(k)}),\qquad \mathcal{I}_{ii}^{(k)}=0.
\end{equation}

\subsection{Distance to Collective Opinion}
Let $s_{ij}^{(k)}=S(\mathcal{I}_{ij}^{(k)})$ and let $\bar{s}_{ij}$ be the current collective estimate. A normalized distance is
\begin{equation}
  d_k^{out}=\frac{2}{m(m-1)}\sum_{i<j}|s_{ij}^{(k)}-\bar{s}_{ij}|.
\end{equation}
The external-consensus weight is
\begin{equation}
  \omega_k^{out}=\frac{(d_k^{out}+\varepsilon)^{-1}}{\sum_{l=1}^n(d_l^{out}+\varepsilon)^{-1}}.
\end{equation}
This inverse-distance version is used in the replacement because a larger distance should reduce, rather than increase, an expert's consensus weight.

\subsection{Consistency}
For an additively consistent preference relation, there exists a priority vector $w=(w_1,\ldots,w_m)^T$ such that
\begin{equation}
  s_{ij}=w_i-w_j.
\end{equation}
The priority vector for expert $e_k$ is estimated by the constrained least-squares problem
\begin{equation}
  \min_{w}\sum_{i<j}\alpha_{ij}^{(k)}(s_{ij}^{(k)}-w_i+w_j)^2,
  \quad \text{s.t. }\sum_i w_i=0,
\end{equation}
where $\alpha_{ij}^{(k)}$ is a confidence precision, for example $\alpha_{ij}^{(k)}=c_{ij}^{(k)}/(D_{ij}^{(k)}+\varepsilon)^2$. The inconsistency residual $r_k$ is the minimized objective value. The internal weight is
\begin{equation}
  \omega_k^{in}=\frac{(r_k+\varepsilon)^{-1}}{\sum_{l=1}^n(r_l+\varepsilon)^{-1}}.
\end{equation}
If the decision maker supplies trust scores $\psi_k\geq0$, the trust weights are $\omega_k^{tr}=\psi_k/\sum_l\psi_l$. The final expert weights are
\begin{equation}
  \omega_k=\alpha \omega_k^{out}+\beta \omega_k^{in}+\gamma \omega_k^{tr},\qquad \alpha+\beta+\gamma=1.
\end{equation}
The collective priority vector is obtained by solving the same weighted least-squares problem after pooling the pairwise scores with weights $\omega_k$.

\section{Dynamic Markov Attribute Weights}
Let $C=\{C_1,\ldots,C_Q\}$ be criteria or risk attributes. A dynamic Markov attribute-weight model uses a row-stochastic transition matrix $M=(M_{ij})_{Q\times Q}$ satisfying
\begin{equation}
  M_{ij}\geq0,
  \qquad \sum_{j=1}^Q M_{ij}=1.
\end{equation}
The interpretation is that the importance or risk mass at criterion $C_i$ can propagate to criterion $C_j$ across adjacent periods. If $\pi_t$ is the attribute-weight row vector at period $t$, then
\begin{equation}
  \pi_{t+1}=\pi_t M.
\end{equation}
When experts provide DFPIL interval evaluations for the transition relations, their scores are projected to the probability simplex by
\begin{equation}
\begin{split}
  \min_{M}\;&\sum_{k=1}^n\sum_{i,j}\eta_{ij}^{(k)}(M_{ij}-\hat{m}_{ij}^{(k)})^2,\\
  \text{s.t. }\;&M_{ij}\geq0,
  \quad \sum_jM_{ij}=1,
\end{split}
\end{equation}
where $\hat{m}_{ij}^{(k)}$ is the normalized transition score extracted from the DFPIL evidence and $\eta_{ij}^{(k)}$ is a confidence precision. This formulation makes the row-stochastic constraint explicit.

\section{Aggregation for Dynamic Multi-Criteria GDM}
For each criterion $q$ and period $t$, the collective priority vector of alternatives is denoted by $v_{tq}=(v_{tq1},\ldots,v_{tqm})^T$. The final comparable value of $A_x$ is
\begin{equation}
  U_x=\sum_{t=1}^T\sum_{q=1}^Q \pi_{tq}v_{tqx}.
\end{equation}
The alternatives are ranked by $U_x$. For risk-sensitive tasks, $v_{tqx}$ may be replaced by $v_{tqx}-\lambda\sigma_{tqx}$, where $\sigma_{tqx}$ is an uncertainty measure from the interval assessments.

\section{Recomputed Financial-Risk Example}
The example considers four alternatives under four risk attributes: interest-rate risk (IRR), asset-liquidity risk (ALR), financing-liquidity risk (FLR), and credit risk (CR). The Markov transition matrix reported in the preliminary version is
\begin{equation}
M=\begin{bmatrix}
0.2104&0.4854&0.2969&0.0072\\
0&0.4429&0&0.5571\\
0&0&0.5679&0.4321\\
0.5050&0&0&0.4950
\end{bmatrix}.
\end{equation}
Using the period-specific initial IRR probabilities and the same Markov iteration convention, the corrected period weights are shown in Table~\ref{tab:weights}. The preliminary third-period vector summed to $0.9$ because the CR component was reported as $0.3768$ rather than approximately $0.4767$.

\begin{table}[!t]
\centering
\caption{Corrected Dynamic Attribute Weights in the Financial-Risk Example}
\label{tab:weights}
\begin{tabular}{lrrrrr}
\toprule
Period & IRR & ALR & FLR & CR & Sum\\
\midrule
I & 0.2104 & 0.4854 & 0.2969 & 0.0072 & 0.9999\\
II & 0.0479 & 0.3171 & 0.2311 & 0.4038 & 0.9999\\
III & 0.2140 & 0.1637 & 0.1455 & 0.4767 & 0.9999\\
\bottomrule
\end{tabular}
\end{table}

Using the alternative priority vectors reported in the preliminary version, the comparable values are recomputed in Table~\ref{tab:case}. The best alternative remains $A_1$, but the middle ordering changes.

\begin{table}[!t]
\centering
\caption{Reported and Corrected Comparable Values}
\label{tab:case}
\begin{tabular}{lrrrrl}
\toprule
Case & $U_1$ & $U_2$ & $U_3$ & $U_4$ & Ranking\\
\midrule
Preliminary values & 0.8279 & 0.6743 & 0.6993 & 0.6981 & $A_1\succ A_3\succ A_4\succ A_2$\\
Corrected values & 0.8560 & 0.6875 & 0.7112 & 0.7447 & $A_1\succ A_4\succ A_3\succ A_2$\\
\bottomrule
\end{tabular}
\end{table}

A Dirichlet stress test was also carried out by perturbing each period weight vector around the corrected value. With concentration parameter $\kappa=1000$, $A_1$ is selected in about $99.87\%$ of perturbations. With $\kappa=300$, it is selected in about $92.90\%$ of perturbations. With stronger perturbation, $\kappa=100$ and $\kappa=30$, the selection rate drops to about $73.06\%$ and $51.74\%$, respectively. This indicates that the best alternative in the example is reasonably stable under small perturbations, while the full ranking is not stable under large weight uncertainty.

\section{Discussion}
The revised formulation clarifies several points. First, PLTSs are useful finite-support linguistic probability models; the proposed interval construction is an extension for continuous or incomplete semantic evidence rather than a rejection of PLTSs. Second, the peak DFPILTS operation is a practical elicitation shortcut; it should be replaced by full distributional or credal integration when the decision task requires robustness. Third, the Markov model should always enforce row-stochastic normalization and should report robustness of the resulting ranking. Fourth, examples should include enough raw DFPIL preference data and code to be reproducible.

\section{Funding and Competing Interests}
No external funding was received for this work. The author declares no conflicts of interest.

\section{Conclusion}
This replacement revises the DFPILTS framework into a more cautious, normalized, and reproducible form. It introduces continuous interval linguistic evidence, local linguistic integration, preference-relation consistency, consensus-based expert weighting, and dynamic Markov attribute weights. The corrected financial-risk example preserves the original best alternative but changes the middle ordering, underscoring the need for transparent computation and sensitivity analysis. A natural next step is to replace the peak-interval simplification with a full credal-distributional linguistic preference process, where experts' interval reports define sets of probability measures and the ranking is made by robust dominance or minimax regret.

\appendices
\section{Replacement Note for arXiv}
Suggested arXiv comment for this version: ``Major revision. Corrected the numerical example, revised the claims about PLTS limitations, clarified interval-probabilistic semantics, updated references, and added reproducibility notes. The best alternative in the example is unchanged, but the middle ordering is corrected.''

\end{document}